\documentclass[12pt]{article}
\usepackage{latexsym,epsfig,graphicx,a4wide,amssymb}

\def\a{\alpha}

{\rm }

\def\be{\begin{equation}}
 \def\ee{\end{equation}}
 \def\bea{\begin{eqnarray}}
 \def\eea{\end{eqnarray}}
 \def\a{\alpha}

\newcommand{\fr}{\frac}
\newcommand{\pr}{\prime}

\def\2{\frac{1}{2}}
\def\4{\frac{1}{4}}

\catcode`\@=11

\@addtoreset{equation}{section}

\def\@normalsize{\@setsize\normalsize{15pt}\xiipt\@xiipt
\abovedisplayskip 14pt plus3pt minus3pt%
\belowdisplayskip \abovedisplayskip
\abovedisplayshortskip  \z@ plus3pt%
\belowdisplayshortskip  7pt plus3.5pt minus0pt}
\def\small{\@setsize\small{13.6pt}\xipt\@xipt
\abovedisplayskip 13pt plus3pt minus3pt%
\belowdisplayskip \abovedisplayskip
\abovedisplayshortskip  \z@ plus3pt%
\belowdisplayshortskip  7pt plus3.5pt minus0pt
\def\@listi{\parsep 4.5pt plus 2pt minus 1pt
            \itemsep \parsep
            \topsep 9pt plus 3pt minus 3pt}}

\def\underline#1{\relax\ifmmode\@@underline#1\else
        $\@@underline{\hbox{#1}}$\relax\fi}
\@twosidetrue \relax

\catcode`@=12

\catcode`\@=11

\def\section{\@startsection{section}{1}{\z@}{3.5ex plus 1ex minus
   .2ex}{2.3ex plus .2ex}{\large\bf}}

\def\ps@headings{\def\@oddfoot{}\def\@evenfoot{}
\def\@oddhead{\hbox{}\hfill
        \makebox[.5\textwidth]{\raggedright\ignorespaces --\thepage{}--
        \hfill }}
\def\@evenhead{\@oddhead}
\def\subsectionmark##1{\markboth{##1}{}}
}

\ps@headings

\catcode`\@=12

\begin{document}

\begin{titlepage}
%
%


%

\begin{centering}
\vspace{1cm}
{\Large {\bf Charged C-metric with conformally coupled scalar
field}}
\\

\vspace{2cm}

 {\bf Christos Charmousis $^{\flat}$} \\
   LPT, CNRS UMR 8627 , Univesit$\acute{e}$ de Paris-Sud, Bat. 210, 91405 Orsay CEDEX,
 France\\ LMPT, CNRS  UMR 6083, Universit\'e Fran\c cois Rabelais - Tours\\
\vspace{.3in}

 {\bf Theodoros Kolyvaris $^{\sharp}$}, {\bf Eleftherios Papantonopoulos $^{*}$} \\
 Department of Physics, National Technical University of
Athens, \\
Zografou Campus GR 157 73, Athens, Greece \\
\vspace{.2in}

\vspace{3mm}

\end{centering}
\vspace{2.5cm}

\begin{abstract}

We present a generalisation of the charged C-metric conformally
coupled with a scalar field in the  presence of a cosmological
constant.  The solution is asymptotically flat or a constant
curvature spacetime. The spacetime metric has the geometry of a
usual charged C-metric with cosmological constant, where the mass
and charge are equal. When the cosmological constant is absent it is
found that the scalar field only blows up at the angular pole of the
event horizon. The presence of the cosmological constant can
generically render the scalar field regular where the metric is
regular, pushing the singularity beyond the event horizon.
For certain cases of enhanced acceleration with a negative cosmological constant, the
conical singularity disappears alltogether and the scalar field is everywhere regular. The black hole is then
rather a black string with its event horizon extending all the way to asymptotic infinity and providing itself the necessary acceleration

\end{abstract}

\vspace{3.5cm}
\begin{flushleft}
$^{\flat}~~$ e-mail address: Christos.Charmousis@th.u-psud.fr \\
$^{\sharp}~~$ e-mail address: teokolyv@central.ntua.gr \\
 $ ^{*} ~~$ e-mail address: lpapa@central.ntua.gr

\end{flushleft}
\end{titlepage}

\section{Introduction}

The C-metric is a static and axially symmetric solution of the
Einstein equations. It has the symmetries of a  Weyl metric (see the
analysis in \cite{gregory} including a cosmological constant) and
belongs to the special Class of Petrov type D metrics just as most
common  black hole solutions of the vacuum. The solution describes
two uniformly accelerated black holes in opposite directions. To
compensate the mutual gravitational attraction of the black holes, a
conical singularity develops in one of its angular poles. This nodal
singularity, often present in multi-black hole Weyl metrics, was
interpreted~\cite{Kinnersley:1970zw} as due to the presence of a
strut in between keeping  the black holes away, or as two strings
from infinity pulling in each one of the black holes in such a way
as to obtain an (unstable) static equilibrium. The strut or the
strings lie along the symmetry axis and they can be seen as the
cause (or the effect) of the uniform acceleration of the black hole
pair. This nodal singularity of a charged C-metric can be removed by
an appropriate transformation introducing an external
electromagnetic field~\cite{Ernst}. In this new exact Ernst C-metric
solution the acceleration of the pair of oppositely charged black
holes is provided by the Lorentz force associated to the external
magnetic field. The geometrical properties and physical
interpretation of the C-metric were further developed and
discussed~\cite{c-metric1}.

The C-metric can be generalised by  introducing into the solution a NUT
parameter, a rotation and a cosmological constant term
$\Lambda$~\cite{PlebDem}, while in~\cite{DGKT}
 a dilaton field non-minimally coupled was  included into
 the solution. A careful and ingenious mathematical reincarnation of this solution led to the
five-dimensional black ring metric found by Emparan and Reall
\cite{empreal} which opened up a whole
 new subject in higher dimensional black holes \cite{emp}. The cosmological constant
version of these higher dimensional black holes has been however obstensively eluding discovery \cite{gregory}, \cite{langlois} (see however the approximative methods of \cite{niar}). The flat spinning C-metric was
 further studied~\cite{FarhZimm4,LetOliv,BicPrav} and in particular in
\cite{BicPrav}, the flat spinning C-metric has been transformed into
the Weyl form and interpreted as two uniformly accelerated spinning
black holes connected by a strut.

The presence of a cosmological constant does not change the generic
features of the C-metric. The cosmological constant length scale, $l$ plays a
complimentary role  to the acceleration parameter $A$ shifting the
horizon positions. In particular, in the case of adS space there is
a relation between the acceleration parameter $A$ and  $l$. The case
$A<1/l$ was studied in~\cite{Pod} while the case $A=1/l$ was
investigated in~\cite{EHM1}. The case $A>1/l$ was extensively
studied in~\cite{Dias:2002mi} where an analysis of the causal
structure of the solution showed that it describes a pair of
accelerated black holes in an adS background.

The Euclidean version of the  C-metric has been used for pair creation of
black holes. In~\cite{DGKT} a generalization of the flat C-metric with a
dilaton field, amended  with a flat Ernst solution, to ensure a regular instanton,  was applied in
the context of quantum pair creation of black holes, that once
created via quantum tunneling accelerate apart. Also the C-metric solution for generic
$\Lambda$ has been used \cite{HHR,MannAdS,Osc} to describe the final
state of the quantum process of pair creation of black holes.  The
quantum process that might create the pair would be the
gravitational analogue of the Schwinger pair production of charged
particles in an external electromagnetic field.

The usual form of the C-metric has a structure function which is a
cubic polynomial in its variable, while in the charged case the structure
function is quadric. A new neat form of the charged C-metric was proposed
in~\cite{Hong:2003gx} in which the structure function after a
coordinate transformation was written in a factorisable form. This
form of the C-metric leads to considerable simplifications, intuitive understanding,  and
allows to cast the metric in Weyl coordinates. An analogous new form
of the charged rotating C-metric was also
proposed~\cite{Hong:2004dm} with a factorisable structure function.
The difference with the non-rotating case is that this new form of
the metric is not related to the old one by a coordinate
transformation.

In this work we will present a generalization of the charged
C-metric with conformally coupled scalar hair. The metric
solution is similar to the $M=Q$ charged C-metric{\footnote{One is
tempted to call this case extremal but here the inner Cauchy and event
horizon do not coincide as for the RN metric.}} but includes a scalar and EM field which
have electric, magnetic and scalar charge linked by one relation to
the mass and cosmological constant. In the limit of zero
acceleration and zero cosmological constant we find the conformally
coupled scalar black hole solution of \cite{BBMB}. This solution has
a regular metric horizon covering a trapped  black hole geometry
with a non-trivial scalar field. The scalar field however,
blows up at the black hole's horizon in accordance to "no hair"
expectations. If one allows for a non-zero cosmological constant one
can push the scalar singularity behind the event horizon,  and
obtain  a regular solution of a charged black hole conformally
coupled to a scalar field as found in~\cite{Martinez:2002ru} and
\cite{Martinez:2004nb}. For the solution we find here, we will see
that acceleration plays a similar role to a cosmological constant.
We will find that for a choice of parameters the scalar field
is regular at the horizon and beyond  if $\Lambda \neq 0$ and it is
singular at the conical tip of the horizon in the flat case. Furthermore, for negative cosmological constant and enhanced acceleration
we will see that the accelerated black hole becomes much like a black string with its event horizon reaching out all the way to asymptotic  infinity. This case  follows closely the description of the constant vacuum adS C-metric  given in \cite{EHM1}. We will see that the scalar is then well behaved in the whole of the permitted coordinate region and furthermore, the metric has no longer a conical singularity. Following \cite{EHM1} we can interpret this as the string being smoothed out  by the accelerating black string itself, extending all the way to asymptotic infinity. In this sense the conical singularity is replaced by the black hole horizon itself! This effect therefore is not due to the scalar field as one could have naively originally thought. We present the solution in the next section and then transform to Hong-Teo \cite{Hong:2003gx} coordinates in order to conclude in the last section.

\section{Accelerating Charged Black Hole Coupled to a Scalar Field}

Consider the action \bea\label{confaction}
    I=\int
    d^4x\sqrt{-g}\left[\frac{R-2\Lambda}{16\pi}-\frac{1}{2}g^{\mu\nu}\partial_\mu\phi\partial_\nu\phi-\frac{1}{12}R\phi^2-\alpha\phi^4-\frac{1}{16\pi}F_{\mu\nu}F^{\mu\nu}\right]~,
\eea where $\alpha$ is a dimensionless constant. The corresponding
field equations are ($G=c=1$) \bea\label{fieldeqs}
    G_{\mu\nu}+\Lambda g_{\mu\nu} &=&8\pi \left( T_{\mu\nu}^{^{(S)}}+T_{\mu\nu}^{^{(EM)}}\right)~, \\
    \square\phi&=&\frac{1}{6}R\phi+4\alpha\phi^3~,\\
    \partial_\nu[\sqrt{-g}F^{\mu\nu}]&=&0~,\label{fieldeqs1}
\eea where the energy-momentum tensors of the scalar and
electromagnetic fields are respectively,
 \bea\label{scalmom}
T_{\mu\nu}^{^{(S)}} & = & \partial_\mu\phi\partial_\nu\phi-\frac{1}{2}g_{\mu\nu}g^{\alpha\beta}\partial_\alpha\phi\partial_\beta\phi+\frac{1}{6}\left[g_{\mu\nu}\square-\nabla_\mu\nabla_\nu+G_{\mu\nu}\right]\phi^2-\alpha g_{\mu\nu}\phi^4~, \\
\label{emmom} T_{\mu\nu}^{^{(EM)}} & = &
\frac{1}{4\pi}g^{\alpha\beta}\left[F_{\mu\alpha}F_{\nu\beta}-\frac{1}{4}g_{\mu\nu}g^{\gamma\delta}F_{\gamma\alpha}F_{\delta\beta}\right]~,
\eea and $\square\equiv g^{\mu\nu}\nabla_{\mu}\nabla_{\nu}$.

The scalar field is non-trivially and conformally coupled, therefore
the total energy-momentum tensor $T_{\mu\nu}$ is traceless, and so
the Ricci scalar curvature is constant, $R=4\Lambda$. In fact, since
$T_{\mu\nu}^{^{(S)}}$ is traceless,  it will play, for particular
couplings, the same role as the EM energy-momentum tensor
$T_{\mu\nu}^{^{(EM)}}$. The field equations
(\ref{fieldeqs})-(\ref{fieldeqs1}) admit the following solution,
\bea\label{confmetric}
    && ds^2=\fr{1}{A^2 (x-y)^2}\left[ F(y) dt^2-\frac{1}{F(y)} dy^2+\fr{1}{G(x)} dx^2+G(x) d\varphi^2 \right]~, \\
    \label{Fofy}
    && F(y) = \frac{\Lambda}{3A^2} + 1 - y^2 - 2mA y^3 - m^2 A^2 y^4~, \\
    \label{Gofx}
    && G(x) =   1 - x^2 - 2mA x^3 - m^2 A^2 x^4~, \\
    && \phi(y , x) = \sqrt{-\frac{\Lambda}{6\alpha}}\:\frac{Am (x-y)}{1+Am(x+y)}~, \\
    \label{em-pot}
    && \mathcal{A} =  e y dt + g x d\varphi~,
\eea where $e $ and $g$  are the electric and magnetic parameters
respectively and  they are related to the mass through the relation
\bea
    e^2 + g^2 = m^2 \left( 1+\fr{2 \pi \Lambda}{9 \a} \right)~.
    \label{ratio}
\eea The geometry of the black hole (\ref{confmetric}) is the same
as the charged C-metric,  with the only difference that the ratio of
the electromagnetic charges to mass has a bound as can be seen in
(\ref{ratio}) and that the usual C-metric polynomials $F$ and $G$
are identical to the charged electric C metric of mass $M$ and
charge $Q$ for $M=Q$. In fact, under $(\ref{ratio})$, it would seem
that the scalar and EM field of each black hole act merely as test
fields in the $Q=M$ black hole geometry, allowing a static
equilibrium to be obtained.

We can also write the solution in the Einstein frame. Permorming a
conformal transformation, with a scalar redefinition of the form
\bea
    \widetilde{g}_{\mu\nu} = \left( 1 - \fr{4\pi}{3} \phi^2\right) g_{\mu\nu} \quad , \quad
    \Psi = \sqrt{\fr{3}{4\pi}}\mbox{Arctanh}\left(
    \sqrt{\fr{4\pi}{3}}\phi\right)~,
\eea the action (\ref{confaction}) becomes
\bea
    I=\int d^4x \sqrt{-\widetilde{g}} \left[\frac{\widetilde{R}-2\Lambda}{16\pi}
-\frac{1}{2}\widetilde{g}^{\mu\nu}\partial_\mu\Psi\partial_\nu\Psi - V(\Psi)  - \frac{1}{16\pi}
    F_{\mu\nu}F^{\mu\nu}
    \right]~,
\eea
where the self-interaction potential is
\bea
\label{teos}
    V(\Psi) = \fr{\Lambda}{8\pi}\left[\cosh^4\left(\sqrt{\fr{4\pi}{3}}\Psi \right)  + \fr{9\a}{2\pi\Lambda}\sinh^4\left(\sqrt{\fr{4\pi}{3}}\Psi \right)
    -1\right]~.
\eea

In this frame the solution takes the form \bea\label{einsteinmetric}
    && ds^2= \fr{u(y,x) }{A^2 (x-y)^2}\left[ F(y) dt^2-\frac{1}{F(y)} dy^2+\fr{1}{G(x)} dx^2+G(x) dz^2 \right]~, \\
    && u(y,x) =1+\fr{2\pi\Lambda}{9\a}\left( \frac{Am (x-y)}{1+Am(x+y)} \right) ^2~, \\
    && \Psi(y , x) =  \sqrt{\fr{3}{4\pi}}\mbox{Arctanh}\left( \sqrt{-\frac{2\pi\Lambda}{9\alpha}}\:\frac{Am (x-y)}{1+Am(x+y)} \right)~,
\eea with $ F(y)$, $ G(x)$ and $\mathcal{A}$ given by (\ref{Fofy}),
(\ref{Gofx})  and (\ref{em-pot}). In this frame the scalar $\Psi$ is
minimally  coupled with the expense of having a precise
self-interaction scalar potential rather than a simple cosmological
constant in the action.  The limit of $\Lambda\rightarrow 0$ is
obtained upon letting the coupling $\alpha\rightarrow 0$ so that
$\frac{\alpha}{\Lambda}\sim constant$.  This solution then reduces
smoothly to the solution found in \cite{BBMB} for zero acceleration
$A=0$. When $A\neq 0$ but $\Lambda=0$ the potential drops out giving
the solution of \cite{DGKT} at minimal EM-scalar coupling.  If
finally on the other hand  $\Lambda\neq 0$, we obtain the solutions
found in \cite{Martinez:2002ru}, \cite{Martinez:2004nb},
\cite{Martinez:2005di}.

\subsection{Properties of the Solution}
The solution we have presented will follow closely the properties of the $Q=M$ charged C-metric, (see for example \cite{Dias:2002mi}). Here, we give the generic properties and we concentrate on the possible regularity of the scalar field. This is in part relevant to the no-hair theorems for general relativity.
In order for the metric (\ref{confmetric}) to have the right signature $
G(x)$ must be positive whereas $F(y)$ has to be negative. For $Am\neq 0$ equation $G(x)=0$ has up to four
real roots \bea
    \xi_1 & = & \fr{-1-\sqrt{1+4Am}}{2Am}~, \\
    \xi_2 & = & \fr{-1-\sqrt{1-4Am}}{2Am}~, \\
    \xi_3 & = & \fr{-1+\sqrt{1-4Am}}{2Am}~, \\
    \xi_4 & = & \fr{-1+\sqrt{1+4Am}}{2Am}~.
\eea
When $\Lambda=0$ we have $F(\xi)=G(\xi)$ and the situation is relatively simple. When $\Lambda\neq 0$ the angular sections $y=constant$ are similar but the $x=constant$ sections are shifted around.
To find the horizons we must solve the equation $F(y)=0$, which
a priori gives four solutions,
 \bea
    y_1 & = & \fr{-1-\sqrt{1+4Am\sqrt{1+\Lambda/3A^2}}}{2Am}~,\label{hor1} \\
    y_2 & = & \fr{-1-\sqrt{1-4Am\sqrt{1+\Lambda/3A^2}}}{2Am}~,  \\
    y_3 & = & \fr{-1+\sqrt{1-4Am\sqrt{1+\Lambda/3A^2}}}{2Am}~,  \\
    y_4 & = & \fr{-1+\sqrt{1+4Am\sqrt{1+\Lambda/3A^2}}}{2Am}~. \label{hor4}
\eea

For the $y_i$ to be real it is necessary to have
$1+\Lambda/3A^2\geq0$ which is always satisfied for $\Lambda\geq
0$ and for  adS space the relation is satisfied if $\mid A\mid ~
>1/l$. Now $y_2,y_3$ are real provided that
$Am\leq\fr{1}{4\sqrt{1+\Lambda/3A^2}}$ and $y_1,y_4$ are real when
$Am\geq-\fr{1}{4\sqrt{1+\Lambda/3A^2}}$.  At $y\rightarrow
-\infty$ we have a curvature singularity, and therefore the usual
radial coordinate is roughly speaking $r \sim -\frac{1}{y}$. Null
and timelike infinity arises at the zero for the conformal factor
at $x=y$. To have $G(x)>0$, we restrict the $x$ coordinate to the
range $(x_s,x_n)$, while $y$ must belong to the range $-\infty\leq
y <x$ for all $x\in (x_s,x_n)$. Depending on the value of $A m$ we
have the following cases:

\begin{center}
\textbf{Case I.} $\mathbf{Am > 0}$
\end{center}
\textbf{Ia.} $\mathbf{Am < \fr{1}{4}}$ \\
\ \\
In this case the equation $G(x)=0$ has four real roots, where we have,
 $\xi_4>0>\xi_3>\xi_2>\xi_1$ and
\begin{eqnarray*}
    G(x)>0\Rightarrow \left\{
        \begin{array}{l}
                \xi_3 < x < \xi_4 \\
                \qquad\textrm{or} \\
            \xi_1 < x < \xi_2
        \end{array} \right.
\end{eqnarray*}

We a priori concentrate on the interval  $\xi_3 \leq x \leq \xi_4 $
for the metric to have Lorentz signature. The axis $ x = \xi_3 $
points towards spatial infinity, and the axis $ x = \xi_4 $ points
towards the other black hole. The surface $ y = y_1 $ is the inner
black hole horizon, $ y = y_2 $ is the event horizon and $ y = y_3 $
the acceleration horizon.  Now, as we noted earlier, for $\Lambda=0$
the roots of $F$ and $G$ coincide, therefore the spacelike $y$
region is $(\xi_2, \xi_3)$ and $y<x$ for all  $\xi_3 \leq x \leq
\xi_4$. Therefore the surfaces of constant $y$ are topological
spheres \cite{horo}. Allowing a positive cosmological constant
shrinks the $(y_2,y_3)$ interval, i.e $\xi_2<y_2<y_3<\xi_3$ and thus
the topology of the angular sections remains spherical. This is to
be expected since a positive cosmological constant "enhances" the
acceleration. In order for the event and acceleration horizon to
exist we have a lowest bound for the black hole mass namely,
$\frac{1}{16m^2}\geq A^2+\frac{\Lambda}{3}$. On the contrary for
$\Lambda<0$ we have  $y_2<\xi_2<\xi_3<y_3$ and whenever $y>\xi_3$ we
only have one axis (since $y<x$) \cite{EHM1}. This means that there
is a topology change as $y>\xi_3$ and the spaces of constant $y$ are
then topologically $ R^2$. The horizon $y=y_3$ is now the adS
horizon.

Let us now examine the behavior of the scalar field in the allowed
range of $x$ and $y$. What we are interested in, is the scalar
field regularity. So we will examine the scalar field denominator. We have
\begin{eqnarray*}
    \Sigma = 1+Am(x+y) \geq 1+Am(\xi_3 + y_2) = \fr{1}{2}\left(\sqrt{1-4Am}-\sqrt{1-4Am\sqrt{1+\fr{\Lambda}{3A^2}}}\right)
\end{eqnarray*}
The locus of $\Sigma=0$ is the region where the scalar field
explodes. We see that for $\Lambda=0$ we have $\Sigma \geq 0$
\footnote{$\Sigma=0$ at the horizon and positive outside, and as
we will see in the next section the singularity is only at the one
pole of $G(x)$.}, for $\Lambda>0$, $\Sigma$ is positive
everywhere, while for $\Lambda<0$, $\Sigma$ can be zero even
before we reach the event
horizon.\\
\\
\textbf{Ib.} $\mathbf{Am \geq \fr{1}{4}}$ \\
\ \\
In this case $\xi_4>0>\xi_1$ and $  G(x)>0 $ is a priori in the
range $\xi_1 < x < \xi_4 $ (for $Am =\frac{1}{4}$ we have $\xi_1 <
\xi_2=\xi_3 < \xi_4 $).  This means that all $\Lambda\geq 0$
solutions are singular since $F<0$ to the left of $y=y_1$ and
therefore we are exposed  to the naked singularity. The only
interesting case is for $\Lambda<0$ where $y=y_2$ and $y=y_3$ can
still be real. Here, we will have $y\leq x< \xi_4$ and spaces of
constant $y$ are topologically $R^2$ planes. In this case,
\cite{EHM1} the event horizon can hit asymptotic infinity so the
solution resembles a black string. However, it is easy to show
that $\Sigma \geq 1+ 2Amy_2$ which always has a solution for
$4Am\geq 1$ and therefore the scalar will be singular for
$y_2<y<y_3$.

\begin{center}
\textbf{Case II.} $\mathbf{Am < 0}$
\end{center}
\textbf{IIa.} $\mathbf{Am > -\fr{1}{4}} $ \\
\ \\
In this case the equation $G(x)=0$ has four real roots, where it
holds: $\xi_2>\xi_1>\xi_4>0>\xi_3$ and
\begin{eqnarray*}
    G(x)>0\Rightarrow \left\{
        \begin{array}{l}
                \xi_3 < x < \xi_4 \\
                \qquad\textrm{or} \\
            \xi_1 < x < \xi_2
        \end{array} \right.
\end{eqnarray*}

We restrict $\xi_1 \leq x \leq \xi_2 $ in order for the metric to
have Lorentz signature and to keep $ y < x $. Null or timelike
infinity is reached when $ y \rightarrow x $. The axis $ x = \xi_1 $
points towards spatial infinity, and the axis $ x = \xi_2 $ points
towards the other black hole. The surface $ y = y_3 $ is the inner
black hole horizon, $ y = y_4 $ is the event horizon and $ y = y_1 $
is now the acceleration horizon.

We have for the denominator of the scalar:
\begin{eqnarray*}
    \Sigma = 1+Am(x+y) \geq 1+Am(\xi_1 + y_4) = \fr{1}{2}\left(\sqrt{1+4Am\sqrt{1+\fr{\Lambda}{3A^2}}}-\sqrt{1+4Am}\right)
\end{eqnarray*}
which is now everywhere positive for $\Lambda<0$, and it can be
zero for $\Lambda>0$, while for $\Lambda=0$ it is zero again on
the horizon. Therefore the situation here  is reversed for the
sign of the cosmological
constant with respect to the case  of $0<Am<\frac{1}{4}$.\\

\textbf{IIb.} $\mathbf{Am \leq -\fr{1}{4}}$ \\

In this case $\xi_2>0>\xi_3$ and $  G(x)>0 $ is a priori in the
range $\xi_3 < x < \xi_2 $ (for $Am =-\frac{1}{4}$ we have $\xi_3
< \xi_4=\xi_1 < \xi_2 $). This means that all $\Lambda\geq 0$
solutions are singular since $F<0$ to the left of $y=y_3$ and
therefore we are exposed  to the naked singularity. The only
interesting case is for $\Lambda<0$ where $y=y_4$ and $y=y_1$ can
still be real. In other words the situation is similar to case Ib.
Here, we will have $y\leq x< \xi_2$ and spaces of constant $y$ are
topologically $R^2$ planes. Once again, \cite{EHM1} the event
horizon reaches all the way up to asymptotic infinity at $y=x$,
hence the solution resembles rather a black string. There is one
important difference from Ia. Here, it is easy to show that
$\Sigma \geq 1+ 2Amy_4=\sqrt{1+4Am\sqrt{1+\fr{\Lambda}{3A^2}}}$
which is always positive (and thus the scalar is always regular)
in the permitted range of the coordinates \bea
    -\fr{1}{4\sqrt{1+\fr{\Lambda}{3A^2}}} < Am \leq
    -\fr{1}{4}~.\label{consin}
\eea
Furthermore, removal of the nodal singularity in the $(x,\varphi)$ sector
implies~\cite{Dias:2002mi} that \bea \label{der-cond}
    G^{\,\pr}(x_s)=-G^{\,\pr}(x_n)~.
\eea We can easily check that the above relation is satisfied if
$(x_s,x_n)=(\xi_2,\xi_3)$ and enhanced acceleration namely, $Am$
is either greater than $1/4$ or less than $-1/4$ (cases Ib, IIb).
It is in particular true in the
  range of  (\ref{consin}). Therefore, the scalar field is regular and there
    are no conical singularities if $Am$ is in the range of (\ref{consin}).  Here we should emphasize that the absence of the
 conical singularity is not a particular feature of the presence of the scalar, rather this is
 true for the $Q=M$ charged C-metric. So how can it be that the black hole can accelerate without a string providing the acceleration? We noted here that the topology of the black hole event horizon changes from spherical to planar. Furthermore we saw that the horizon reaches all the way out to asymptotic infinity. Therefore the string providing the acceleration has been replaced by the black hole-string itself.  The horizon is actually the thickened out regular string itself! The case (\ref{consin}) is the only case where, the metric and scalar field have $C^2$ regularity in the permitted coordinate region.

\section{Factorisable Form of the Solution}
To have a better understanding of the parameters involved, we can
write the metric (\ref{confmetric}) in a more familiar form.  To
do so, it is more convenient first to bring the  function $G(x)$
in a factorisable form. Following~\cite{Hong:2003gx}  we  write
\bea
    G(x)=(1-x^2)(1+A k_1 x)(1+A k_2 x)~,\label{gfunct}
\eea and we make the following coordinate transformations \bea
    x &\rightarrow& B c_0 (x-c_1)~, \\
    y &\rightarrow& B c_0 (y-c_1)~, \\
    t &\rightarrow& \fr{c_0}{B}t~, \\
    \varphi &\rightarrow& \fr{c_0}{B}\varphi~,
\eea where $c_0,c_1,B $ are real constants. In order to preserve the
form of the line element we must have \bea
    A &\rightarrow& \fr{A}{B}~,\nonumber \\
    G(\xi) &\rightarrow& B^2 G(\xi)~.\label{Gtrans}
\eea

From (\ref{Gtrans}) we get: \bea
    m &=& \fr{M}{c_0^2\sqrt{1+2AM}}~, \\
    c_0 &=& \fr{1+AM+2AMc_1}{\sqrt{1+2AM}}~, \\
    c_1 &=& \fr{-1-AM+\sqrt{1+2AM+5A^2 M^2}}{2AM}~, \\
    B   &=& \sqrt{\fr{1+2AM}{1+2AM+A^2 M^2}}~, \\
    k_2 &=& \fr{M}{1+2 A M}~,
\eea and we have set $k_1=M$.

Then, the function $G(x)$ of (\ref{gfunct}) is written \bea
    G(x)=(1-x^2)(1+A M x)\left(1+\fr{AM}{1+2 A M}\:x\right)~,
\eea
with roots \bea
    \xi_1 =1 \quad,\quad \xi_2 =-1 \quad,\quad \xi_3 =-\fr{1}{AM} \quad,\quad \xi_4
    =-\fr{1+2AM}{AM}~.
\eea Suppose $\xi_1>\xi_2>\xi_3>\xi_4$ \footnote{ That is $0 <
\fr{AM}{1+2AM} < AM < 1$}. Then $G(x)>0$ in the range $-1<x<1$ (or
in $\xi_4<x<\xi_3$). We can not remove both conical singularities at
$x=-1$ and $x=+1$, at the same time, since we can easily check that
the condition $G^{\,\pr}(\xi_1)=-G^{\,\pr}(\xi_2)$ does not hold.

To bring the metric in a familiar form we can now set \bea
    y & = & -\fr{1}{A r}~, \\
    x & = & \cos\theta~.
\eea

Then, the solution is written in the form
 \bea\label{HongTeometric}
    && ds^2=\fr{1}{\left(1+A\,r\cos{\theta}\right)^2}\left[ -\fr{f(r)}{r^2} dt^2+\frac{r^2}{f(r)} dr^2+\fr{r^2}{p(\theta)} d\theta^2+r^2\sin^2{\theta} p(\theta) d\varphi^2 \right] \\
    \label{fofr}
    && f(r) = -\frac{\Lambda}{3}r^4 + \left(1-A^2r^2\right)\left(r-M\right)\left(r-\fr{M}{1+2AM}\right)~, \\
    \label{poftheta}
    && p(\theta) =   \left(1+A M\cos{\theta}\right)\left(1+\fr{AM}{1+2AM} \cos{\theta} \right)~, \\
    && \phi(r ,\theta) = \sqrt{-\frac{\Lambda}{6\alpha}}\:\frac{k(Ar\cos{\theta}+1)}{r+k(Ar\cos{\theta}-1)}~, \\
    \label{elemag-pot}
    && \mathcal{A} = - \fr{e}{r} dt + g \cos{\theta}d\varphi~,
\eea where \bea
    k=\fr{M}{1+AM}~.
\eea The electric and magnetic parameters are connected with the
mass through the relation \bea
    e^2 + g^2 = \fr{M^2}{1+2 A M} \left( 1+\fr{2 \pi \Lambda}{9 \a}
    \right)~.
\eea

The constants $A, M, e, g$ represent the acceleration, mass,
electric and magnetic charge of the solution respectively, as we can
see by taking the appropriate limits.

For $A=0$ and $g=0$, $e=0$ (\emph{i.e.} $\a=-2\pi\Lambda/9$) we
find the solution of a four-dimensional black hole in dS space
coupled with a scalar field found in~\cite{Martinez:2002ru}, so
$M$ is the mass of the non-accelerating black hole. For $A=0$,
$g=0$ we find the solution of a charged black hole coupled with a
scalar field~\cite{Martinez:2005di}, so $e$ is the electric
charge, while if we also have $\Lambda=0$ we find the solution
of~\cite{BBMB}. Finally, if there is no scalar field we have the
charged C-metric with $e=M$ (\cite{Kinnersley:1970zw},
\cite{Dias:2002mi},
 \cite{Hong:2003gx}).

 It is interesting to consider the case $\Lambda=e=g=0$. Then the scalar field reads \bea
    \phi(r ,\theta) & = & \sqrt{\frac{3}{4\pi}}\:
    \frac{k(Ar\cos{\theta}+1)}{r+k(Ar\cos{\theta}-1)}~.
\eea

Remember that in the BBMB black hole \cite{BBMB} the scalar field
blows up at the horizon. In our case for $r=r_+=M$ we get \bea
    \phi(r_+,\theta) & = & \sqrt{\frac{3}{4\pi}}\:\frac{M(AM\cos{\theta}+1)}{M(1+AM)+M(AM\cos{\theta}-1)}~, \\
    \Rightarrow \phi(r_+,\theta) & = &
    \sqrt{\frac{3}{4\pi}}\:\frac{AM\cos{\theta}+1}{AM(1+\cos{\theta})}~.
\eea

So the scalar is regular at the horizon except at the point
$\theta=\pi$, \emph{i.e.} at the pole $x=-1$. It would have been
desirable to remove both conical singularities with an external
electric field, employing an  Ernst-like mechanism. In this case,
the scalar field would be regular everywhere.

Following the same procedure as before, we can bring the metric
(\ref{HongTeometric}) in hyperbolic form. Define
 \bea
    y & = & -\fr{1}{A r}~, \\
    x & = & \cosh\theta~.
\eea

the solution can be written in the form
\bea\label{HongTeometric2}
    && ds^2=\fr{1}{\left(1+A\,r\cosh{\theta}\right)^2}\left[ -\fr{f(r)}{r^2} dt^2+\frac{r^2}{f(r)} dr^2+\fr{r^2}{p(\theta)} d\theta^2+r^2\sinh^2{\theta} p(\theta) d\varphi^2 \right] \\
    \label{fofr2}
    && f(r) = -\frac{\Lambda}{3}r^4 - \left(1-A^2r^2\right)\left(r+M\right)\left(r+\fr{M}{1+2AM}\right)~, \\
    \label{poftheta2}
    && p(\theta) =   \left(1-A M\cosh{\theta}\right)\left(1-\fr{AM}{1+2AM} \cosh{\theta} \right)~, \\
    && \phi(r ,\theta) = \sqrt{-\frac{\Lambda}{6\alpha}}\:\frac{k(Ar\cosh{\theta}+1)}{r-k(Ar\cosh{\theta}-1)}~, \\
    \label{elemag-pot2}
    && \mathcal{A} = - \fr{e}{r} dt + g \cosh{\theta}d\varphi~,
\eea where \bea
    k=\fr{M}{1+AM}~,
\eea and the electric and magnetic charge are connected with the
mass through the relation \bea
    e^2 + g^2 = -\fr{M^2}{1+2 A M} \left( 1+\fr{2 \pi \Lambda}{9 \a} \right) .
\eea For $A=0$ and $g=e=0$ (\emph{i.e.} $\a=-2\pi\Lambda/9$) we
obtain the solution found in \cite{Martinez:2004nb}.

\section{Conclusion}

We presented a novel  solution of a charged C-metric coupled to a
scalar field in the presence of a cosmological constant. The
metric geometry is identical  to the "extremal" charged C-metric
the difference being that the charge to mass ratio is bounded. The
solution can be asymptopically dS, adS and flat depending on the
value of the cosmological constant. There is a range of
parameters, where the scalar field is regular on the horizon and
beyond. In particular for $\Lambda=0$ we see that the scalar field
blows up only at the pole of the horizon since acceleration hides
the remaining singular surface. In a sense the proper acceleration
of the black hole plays a similar role as the cosmological
constant but not quite enough to allow scalar hair! For a choice
of parameters, the conical singularities at both poles are absent
similarily to  the $Q=M$ limit of the charged C-metric
\cite{Kinnersley:1970zw}, \cite{Osc}. We saw that in this case the
adS black hole resembles a black string with the event horizon
reaching out to spatial infinity and being of planar topology. In
one case, IIb, the scalar field is both regular and the metric is
free of conical singularities. The black hole is being pulled away
by its proper black string which provides a regularisation of the
nodal singularity. This situation is similar to the one described
in \cite{EHM1} for the adS C-metric.

We have seen that a conformally coupled scalar field plays a very similar role as an EM energy momentum tensor as long as $Q=M$.
It would be interesting to promote this observation to a genuine solution generating method for Weyl or Papapetrou metrics with the inevitable aim to produce rotating black holes with scalar charge.

\section*{Acknowledgments}

It is with great pleasure that we thank Roberto Emparan for
interesting and helpful discussions on the subject of the C-metric.
T.K and E.P. thank the Laboratory of Theoretical Physics of Orsay
for hospitality where part of this work was carried out. Work
supported by the NTUA research program PEVE07 and the PNCG. E.P. is
partially supported by the European Union through the Marie Curie
Research and Training Network UniverseNet (MRTN-CT-2006-035863).
\\
\\
\\
{\bf {Note added in proof:}} When this work was in its final
stage, we got informed that another group is working on a similar
problem. We had email exchanges, and they told us that they were
studying a general case including rotations. In their submitted
paper~\cite{Anabalon:2009qt}, they studied only the static limit
of their metric (eqs (2.3)-(2.6)) which coincides with ours (eqs
(2.7)-(2.11)).

\end{document}